 \documentclass[aps,pra,superscriptaddress,amsmath,amssymb,preprintnumbers,twocolumn,floatfix,showpacs,showkeys,10pt]{revtex4-1}
 \usepackage{amssymb} \usepackage{epsfig}
 \begin{document}
  \title{Distillability sudden death and sudden birth in a two-qutrit system under decoherence of finite temperature  }

\author{You-neng Guo}
\affiliation{ Department of Electronic and Communication Engineering, Changsha University, Changsha, Hunan
410003, People's Republic of China}
\email{guoxuyan2007@163.com}
\author{Mao-fa Fang}
\email{mffang@hunnu.edu.cn}
\affiliation{ Key Laboratory of Low-Dimensional Quantum Structures and
Quantum Control of Ministry of Education, and Department of Physics,
Hunan Normal University, Changsha 410081, People's Republic of
China}
\author{Guo-you Wang}
\affiliation{College of Science, Hunan University of Technology, Zhuzhou 412008, People's Republic of China}
 \author{Ke Zeng}
  \affiliation{ Department of Electronic and Communication Engineering, Changsha University, Changsha, Hunan
410003, People's Republic of China}
 \begin{abstract}
Distillability sudden death and sudden birth in a two-qutrit system under decoherence of finite temperature are studied in detail. By using of the negativity and realignment criterion, it is shown that
certain initial prepared free entangled states may become bound entangled or separable states in a finite time. Moreover, initial prepared bound entangled or separable states also may become distillable entangled states in a finite time.
 \end{abstract}

 \maketitle
\section{Introduction}
As is well-known, a real-word quantum system is unavoidably interacting with its environments, which give rise to quantum system's decoherence, dissipation and dephasing to degrade the coherent of the quantum system. It was initially proposed by Yu and Eberly that,  the entanglement of quantum systems will disentangle in a finite time owe to the interaction between quantum system and its environments, the phenomenon of finite time disentanglement, also was named entanglement sudden death (ESD) [1, 2]. Later this phenomenon was extended to bipartite or multipartite systems interacting with different kinds of noisy environments [3-13].

In the present paper, we have studied the phenomenon of distillability sudden death and sudden birth in a two-qutrit system under decoherence of finite temperature, where the two qutrits are initially prepared in a specific family of Horodecki states.
 For qutrit-qutrit systems, analogous to the definition of ESD, if an initial free entangled state becomes bound entangled or separable states in a finite time under the influence of local decoherence, then we say that it undergoes distillability sudden death [14]. To the contrary, if an initial prepared in bound entangled or separable states becomes distillable entangled states, then it is said that it undergoes distillability sudden birth [15]. Recently great attention has been paid to the development of work on distillability sudden death [16-23] and sudden birth [24-26]
addressing the practical problems of environmental decoherence. In particular, under certain types of environment interaction, certain distillable entangled states may be converted into bound entangled or separable states in a finite time and vice versa. More recently, Mazhar Ali demonstrated
free entangled states may be converted into bound entangled states or separable states in the presence of both collective and multi-local noises [16, 17]. Later this phenomenon was extended to two-qutrit interacting with thermal reservoirs [18, 19] and the presence of the external magnetic field [20, 21]. Besides, Derkacz and  Jak\'{o}bczyk demonstrated the sudden birth of distillable entanglement for two identical three-level atoms coupled to the common vacuum [24-26].

Recently a transformation between bound entanglement and free entanglement has been demonstrated under differen decoherence channels[27-29]. With these previous studies in mind, we naturally ask the question does there exist distillability sudden birth after a specific family of Horodecki states undergoes distillability sudden death$?$ In this paper we answer this question and demonstrate the phenomenon of distillability sudden death and sudden birth in a two-qutrit system under decoherence channel.  The main idea of this paper is to extend the results obtained in Ref. [30], where the dynamics and protection of quantum correlations of a qubit-qutrit system under local amplitude damping channels with finite temperature have been studied.  By using negativity and bound entanglement defined with realignment criterion, distillability sudden death and sudden birth in a two-qutrit system under decoherence of finite temperature are studied in detail. The results show that
certain initial prepared free entangled states may become bound entangled or separable states in a finite time. Moreover, initial prepared bound entangled or separable states also may become distillable entangled states in a finite time.

The paper is organized as follows. In Sec. \textrm{II}, we illustrate the physical model of two-qutrit system under local amplitude damping channels with finite temperature. We briefly review the quantification and characterization of the entanglement and demonstrate the distillability sudden death and sudden birth in a two-qutrit system under local amplitude damping channels with finite temperature in Sec. \textrm{III}. Finally, we give the conclusion in Sec. \textrm{IV}.

 \section{Dynamics of a two-qutrit system under local amplitude damping channels with finite temperature }\label{model}
In this paper, we consider our system is composed of a two-qutrit system $A$ and $B$ under local amplitude damping channels with finite temperature. For qutrits, we consider a three-level system of V-configuration in this paper. We note the
lower level as $|0\rangle$ and two nondegenerate excited levels as $|1\rangle$ and $|2\rangle$.  At finite temperature, we suppose the probability of losing the excitation is $r$, as well as the probability of absorbing the excitation is $1-r$. Then,  the amplitude damping channel at finite temperature can be obtained as [30]
\begin{equation}E_{1}=\sqrt{r} \left(
\begin{array}{ c c c c l r }
1 & 0 & 0 \\
0 &\sqrt{1-p_{1}}& 0  \\
0 &0& \sqrt{1-p_{2}}\\
\end{array}
\right),
\end{equation}
\begin{equation}E_{2}=\sqrt{r} \left(
\begin{array}{ c c c c l r }
0 & \sqrt{p_{1}} & 0 \\
0 &0& 0  \\
0 &0& 0\\
\end{array}
\right),
\end{equation}
\begin{equation}E_{3}=\sqrt{r} \left(
\begin{array}{ c c c c l r }
0 & 0 & \sqrt{p_{2}} \\
0 &0& 0  \\
0 &0& 0\\
\end{array}
\right),
\end{equation}
\begin{equation}E_{4}=\sqrt{1-r} \left(
\begin{array}{ c c c c l r }
\sqrt{1-p_{1}-p_{2}} & 0 & 0 \\
0 &1& 0  \\
0 &0& 1\\
\end{array}
\right),
\end{equation}
\begin{equation}E_{5}=\sqrt{1-r} \left(
\begin{array}{ c c c c l r }
0 & 0 & 0 \\
\sqrt{p_{1}} &0& 0  \\
0 &0& 0\\
\end{array}
\right),
\end{equation}
\begin{equation}E_{6}=\sqrt{1-r} \left(
\begin{array}{ c c c c l r }
0 & 0 & 0 \\
0 &0& 0  \\
\sqrt{p_{2}} &0& 0\\
\end{array}
\right),
\end{equation}
where $p_{1}$ is the decoherence constant induced by spontaneous emission of the
excited excited state$|1\rangle$to the ground
state$|0\rangle$, and $p_{2}$ represents the decoherence constant of the excited state$|2\rangle$to the ground
state$|0\rangle$, respectively. To simplify, we consider that the decoherence decay $p_{i}=p=1-e_{}^{-2 \gamma \tau }$, and $\gamma$ is the spontaneous emission rate.

In order to investigate distillability sudden death and sudden birth in a two-qutrit system under decoherence of finite temperature, we make each qutrit locally interact with two independent amplitude damping channels, respectively. According to the techniques
of Kraus,
we can obtain the elements of the reduced density
matrix as follows:
\begin{eqnarray}
\rho_{AB}(t)=\Sigma_{i=1}^{6}\Sigma_{j=1}^{6}E_{j}^{B}E_{i}^{A}\rho_{AB}(0)E_{i}^{A\dag}E_{j}^{B\dag}.
\end{eqnarray}

In this paper, we only consider a particular initial state [31] 
\begin{eqnarray}
\rho_{\alpha}(0)=\frac{2}{7}|\psi_{+}\rangle \langle \psi_{+}|+\frac{\alpha}{7}\sigma_{+}+\frac{5-\alpha}{7}\sigma_{-},
\end{eqnarray}
where $2 \leq\alpha \leq5$. In Eq. (8) the maximally entangled state $|\psi_{+}\rangle=\frac{1}{\sqrt{3}}(|01\rangle+|10\rangle+|22\rangle)$ is mixed with separable state $\sigma_{+}=\frac{1}{3}(|00\rangle \langle 00|+|12\rangle \langle 12|+|21\rangle \langle 21|)$ and $\sigma_{-}=\frac{1}{3}(|11\rangle \langle 11|+|20\rangle \langle 20|+|02\rangle \langle 02|)$. It was shown that $\rho_{\alpha}(0)$ is separable for $2 \leq\alpha \leq3$, bound entangled for$3 < \alpha \leq4$, and free entangled for $4 < \alpha \leq5$.

\section{Distillability sudden death and sudden birth in a two-qutrit system}  
Before investigate the time-evolved density matrix $\rho_{\alpha}(0)$ maybe undergoes the phenomenon of distillability sudden death and sudden birth in a two-qutrit system under decoherence of finite temperature, we review the quantification and characterization of the entanglement states including free entangled states and bound entangled states. As we know, free entangled states can be well quantified by the negativity.
However, bound entangled states cannot be quantified and detected by the negativity. One usually takes the realignment criterion [32-34] as such a measure, which can detect certain
bound entangled states as well as the quantification of entanglement for qutrit-qutrit systems. The realignment criterion for a given density matrix $\rho$ is defined as
\begin{eqnarray}
R(\rho)=||\rho_{}^{R}||-1,
\end{eqnarray}
where  $\rho_{ij,kl}^{R}=\rho_{ik,jl}$ and the trace norm of $||\rho_{}^{R}||$ is equal to the sum of the absolute values of the eigenvalues of $\rho_{}^{R}$. For a separable state
$\rho$, the realignment criterion implies that $R(\rho)\leq 0$. For a positive partial transpose (PPT) state, the positive value of the quantity $R(\rho)$ can prove the
bound entangled.

\begin{figure}[htpb]
  \begin{center}
     \includegraphics[width=7.5cm]{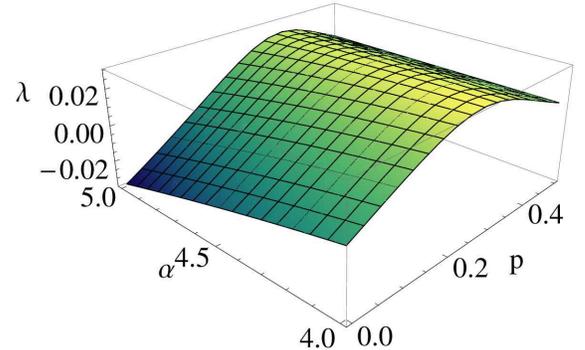}
           \caption{\label{fig1} (Color online) The negative eigenvalue $\lambda(t)$ of the partial transpose of the time-evolved density matrix $\rho(t)$ is plotted against $p$ and $\alpha$, for $r=0.15$.}
  \end{center}
\end{figure}
\begin{figure}[htpb]
  \begin{center}
     \includegraphics[width=7.5cm]{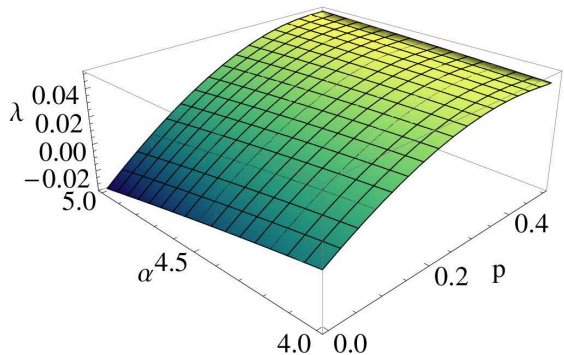}
           \caption{\label{fig2} (Color online) The negative eigenvalue $\lambda(t)$ of the partial transpose of the time-evolved density matrix $\rho(t)$ is plotted against $p$ and $\alpha$, for $r=0.90$.}
  \end{center}
\end{figure}
Besides, to quantify the time evolution of  entanglement for qutrit-qutrit systems, we adopt negativity [35] to
measure of entanglement. It is based on the trace norm of the partial transpose $\rho_{}^{T}$ of the state $\rho$, and its definition is given
\begin{eqnarray}
N(\rho)=||\rho_{}^{T}||-1,
\end{eqnarray}
where  $\rho_{jk,il}^{R}=\rho_{ik,jl}$.
$N(\rho)$ is equal to the absolute of the sum of negative eigenvalues of $\rho_{}^{T}$. For negativity, if it is zero, then we say this quantum state is bound state or separable state. But if it is positive, which means it is negative under partial transposition (NPT), we can not say it must be free entangled because it is still an open question about whether all NPT quantum states can be distilled. For a PPT state which has zero negativity, we cannot conclude its entanglement or separability until some other measures or steps reveal its status. Note that the realignment criterion also cannot detect all bound entangled states. Once the negativity becomes zero, we can study the time evolution of a realignment criterion to detect the possibility of bound entangled states. For an initial state given by Eq. (8) with $4 < \alpha \leq5$, if the norm of the realigned matrix $R(\rho)$ is smaller than 0 corresponding to bound entangled state or separable state, then we say this state maybe suffers from distillability sudden death.
Following we adopt the realignment criterion to investigate the phenomenon of distillability sudden death and sudden birth for two-qutrit system under local amplitude damping channels with finite temperature.
\begin{figure}[htpb]
  \begin{center}
     \includegraphics[width=7.5cm]{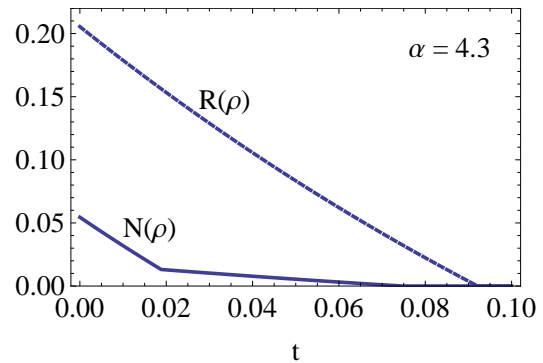}
           \caption{\label{fig3} (Color online) The negativity $N(\rho)$ and the realignment criterion $R(\rho)$ are plotted against $t$ for $\alpha=4.3$, $r=0.90$. Setting $t=\gamma\tau$. }
  \end{center}
\end{figure}
\begin{figure}[htpb]
  \begin{center}
     \includegraphics[width=7.5cm]{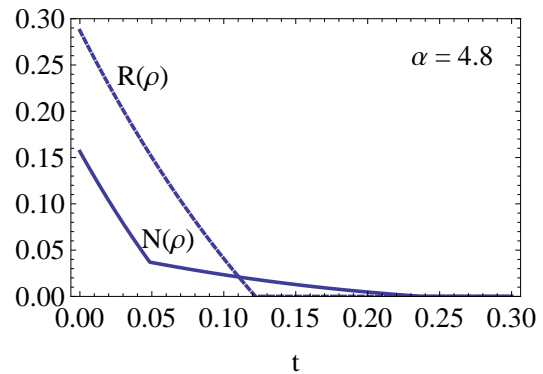}
           \caption{\label{fig4} (Color online) The negativity $N(\rho)$ and the realignment criterion $R(\rho)$ are plotted against $t$ for $\alpha=4.8$, $r=0.90$. Setting $t=\gamma\tau$.}
  \end{center}
\end{figure}

First of all, we analyze the time-evolved matrix density $\rho_{\alpha}(t)$ carefully under local amplitude damping channels with finite temperature. There are three possible negative eigenvalues $\lambda(t)=\lambda_{1}\leq \lambda_{2} \leq \lambda_{3}$ of the partial transposition of the matrix density $\rho_{\alpha}(t)$, we here consider the minimum negative eigenvalue $\lambda(t)=\lambda_{1}$. In Fig. 1 and 2, we plot $\lambda(t)$ as a function of $t$ and $\alpha$  under local amplitude damping channels with finite temperature for $r=0.15$ and $r=0.90$, respectively. We can see that the eigenvalue of partial transposition of the states $\rho(t)$ will always arrive at a positive value in a finite time. This indicates the time-evolved matrix density $\rho_{\alpha}(t)$  always becomes a PPT in a finite time.

Now let us investigate that the time-evolved density matrix $\rho_{\alpha}(0)$ maybe undergoes distillability sudden death. Taking initial states given by Eq. (8) with $4 < \alpha \leq5$. In Fig. 3, we plot the negativity
and realignment criterion against the time $t$ and a specific choice of the single
parameter $\alpha=4.3$ and $r=0.90$, under local amplitude damping channels with finite temperature. It clearly shows that, for a two-qutrit system under local amplitude damping channels with finite temperature, an initial free entangled state becomes bound entangled at a finite time $t\approx 0.075$. This phenomenon is called as distillability sudden death. One can see the positive value of $R(\rho)$ is in the range $0.075\leq t\leq 0.093$ corresponding to the entanglement of the PPT state.  However, it is worth noting that, this realignment criterion fails to detect the possible entanglement after time $t \approx0.093$. This result agrees with previous studies of similar nature for qutrit-qutrit systems [14, 16-21].

\begin{figure}[htpb]
  \begin{center}
  \includegraphics[width=7.5cm]{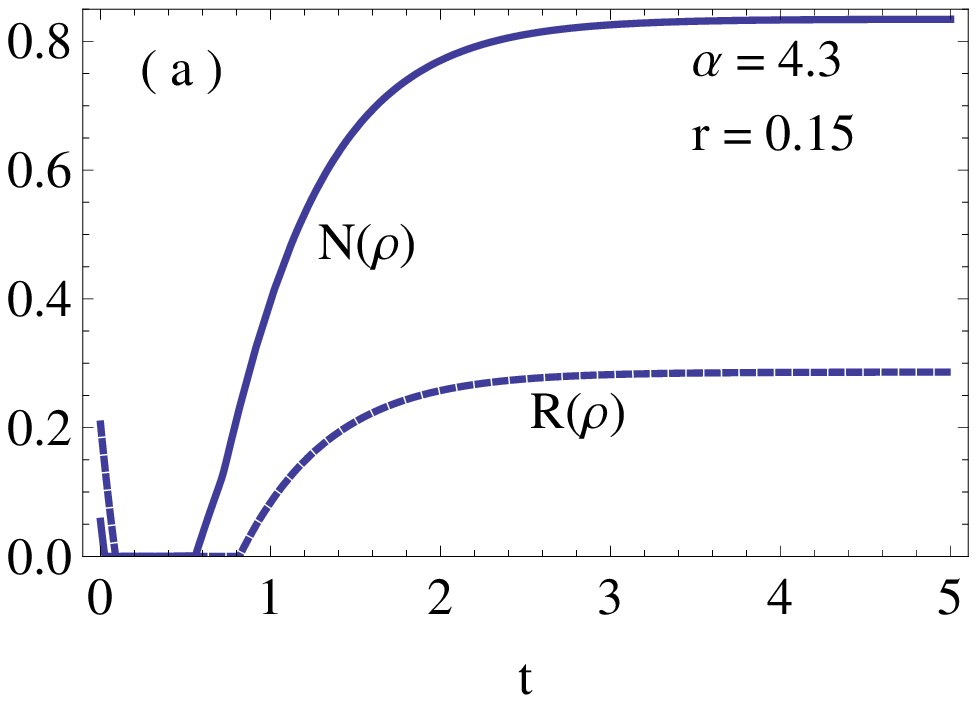}
     \includegraphics[width=7.5cm]{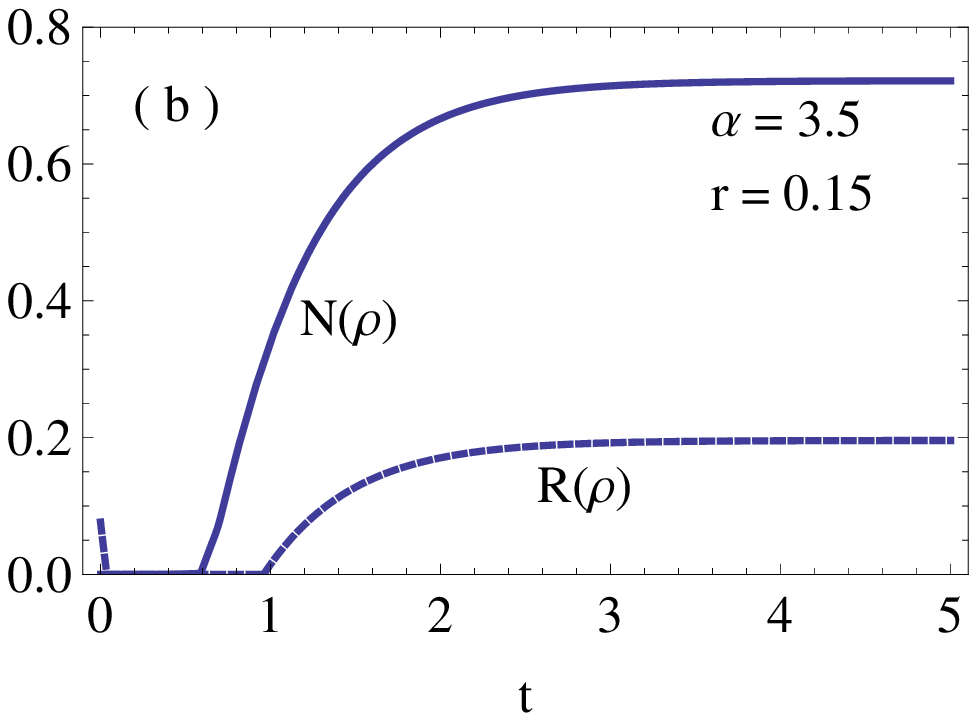}
     \includegraphics[width=7.5cm]{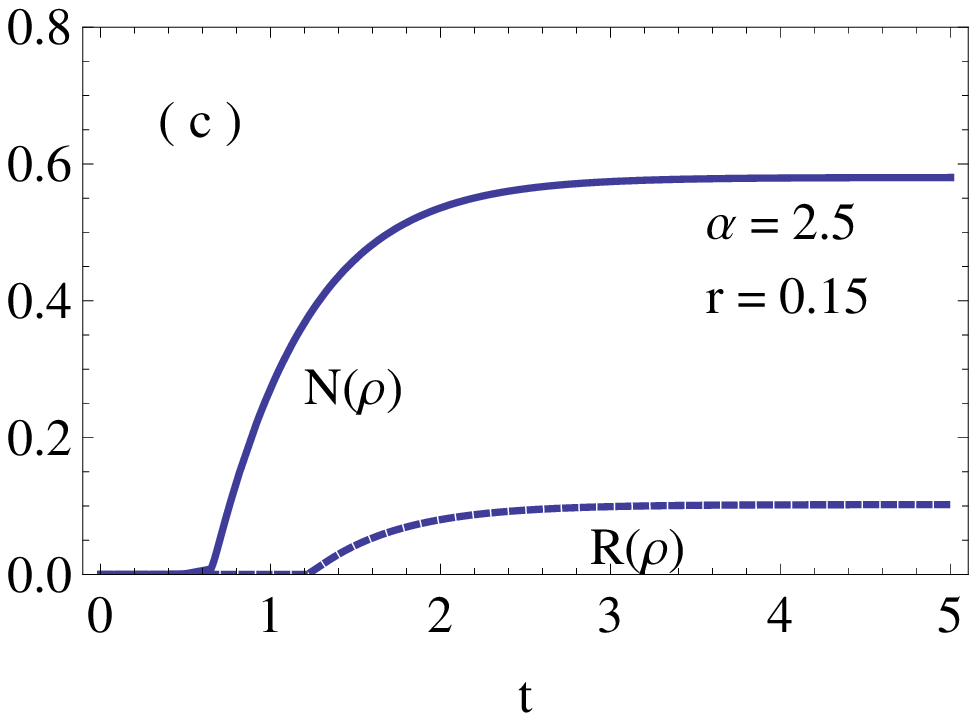}
           \caption{\label{fig5} (Color online) The negativity $N(\rho)$ and the realignment criterion $R(\rho)$ are plotted against $t$ for $\alpha$,  $r=0.15$. Setting $t=\gamma\tau$}
  \end{center}
\end{figure}

In Fig. 4, it is shown the time evolution of the negativity
and realignment criterion for an initial quantum state $\rho_{\alpha}(0)$ with $\alpha=4.8$ and $r=0.90$, under local amplitude damping channels with finite temperature. One can see the negativity becomes zero at $t\approx 0.23$, while the $R(\rho)$ becomes zero at $t\approx 0.12$. As mentioned above, the realignment criterion also cannot detect all bound entangled states, the realignment criterion also fails to detect the bound entangled states after time $t\approx 0.12$.  This means, for this particular case $\alpha=4.8$ as shown in Fig. 4, even if we can see
that the initial NPT states become PPT after a finite time, we can not conclude their
separability or entanglement immediately. By comparing Fig.3 with Fig.4, we find the sensitivity
of the realignment criterion is depended on the initial state parameter $\alpha$ which effects on the possibility of distillability sudden death.

In order to show the distillability sudden death and sudden birth in a two-qutrit system under decoherence induced by spontaneous emission at finite temperature, we firstly investigate initial prepared free entangled states $\rho_{\alpha}(t)$ ($4 < \alpha \leq 5$). In Fig. 5(a), we plot the negativity
and realignment criterion against the time $t$ and a specific choice of the single
parameter $\alpha=4.3$ for $r=0.15$, under local amplitude damping channels with finite temperature. One can see the negativity becomes zero at $0.01\leq t\leq 0.59$, while the $R(\rho)$ becomes zero at $0.05\leq t\leq 0.90$. This means that a two-qutrit system under decoherence induced by spontaneous emission at finite temperature exhibits distillability sudden birth and after distillability sudden death. This result agrees with previous studies of distillability sudden birth of entanglement for qutrit-qutrit systems [15].
However, we are interested in the dynamical creation of distillability by spontaneous emission from non-distillability.
Considering initial prepared bound entangled or separable states $\rho_{\alpha}(t)$ ($2 < \alpha \leq 4$),  we plot the negativity
and realignment criterion against the time $t$ and a specific choice of the single
parameter $\alpha=3.5$ (bound entangled states) in Fig. 5(b) and $\alpha=2.5$ (separable states) in Fig. 5(c) for $r=0.15$, under local amplitude damping channels at finite temperature. Clearly initial prepared bound entangled or separable states  become distillable entangled states in a finite time.
As mentioned above, one can prove one of the eigenvalues of partial transposition of the states $\rho(t)$ can arrive at a negative value again in a finite time. This indicates the time-evolved matrix density $\rho_{\alpha}(t)$  always becomes a NPT in a finite time.
\section{Conclusion} \label{conclusions}

Different from these previous works[14, 16] where are mainly concentrated on distillability sudden death in qutrit-qutrit systems under amplitude damping with zero-temperature, we here have studied the phenomenon of distillability sudden death and sudden birth in a two-qutrit system under decoherence of finite temperature, in which the initial states are prepared in special states.  By using the realignment criterion and negativity, once the negativity becomes zero, we can study the time evolution of a realignment criterion to detect the possibility of bound entangled states. Besides, we have shown initial prepared bound entangled or separable states also may become distillable entangled states in a finite time. However, it is worth pointing that, the possibility of distillability sudden death for two-qutrit system under decoherence of finite temperature is investigated for the family of special quantum states and we do not study this phenomenon in quantum states other than states.

\acknowledgments
This work is supported by the National Natural Science Foundation of China (Grant Nos. 11374096 and 11074072) and Scientific Research Foundation of Hunan Provincial Education Department (No. 13C039).

\label{app:eff-trans}

\end{document}